\documentclass[12pt]{iopart}

\usepackage{graphicx}
\usepackage{hyperref} 
\usepackage{epstopdf}
\usepackage[normalem]{ulem}
\usepackage{color}
\usepackage[utf8]{inputenc}
\usepackage{float}
\usepackage{caption}
\usepackage{bm}
\usepackage{comment}
\usepackage{breqn}
\usepackage{amssymb}
\usepackage{bbold}

\renewcommand\vec[1]{\ensuremath\boldsymbol{#1}}
\begin{document}
\title{Emergence of junction dynamics in a strongly interacting Bose mixture}

\author{R E Barfknecht$^{1,2}$, A Foerster$^2$
and N T Zinner$^{1,3}$}
\address{$^1$ Department of Physics and Astronomy, Aarhus University, Ny Munkegade 120, Denmark}
\address{$^2$ Instituto de F{\'i}sica, Universidade Federal do Rio Grande do Sul, Av. Bento Gon{\c c}alves 9500, Porto Alegre, RS, Brazil}
\address{$^3$ Aarhus Institute of Advanced Studies, Aarhus University, DK-8000 Aarhus C, Denmark}
\eads{\mailto{rafael@phys.au.dk}, \mailto{rafael.barfknecht@ufrgs.br}}

\date{\today}
\begin{abstract} We study the dynamics of a one-dimensional system composed of a bosonic background and one impurity in single- and double-well trapping geometries. In the limit of strong interactions, this system can be modeled by a spin chain where the exchange coefficients are determined by the geometry of the trap. We observe non-trivial dynamics when the repulsion between the impurity and the background is dominant. In this regime, the system exhibits oscillations that resemble the dynamics of a Josephson junction. Furthermore, the double-well geometry allows for an enhancement in the tunneling as compared to the single-well case.
\noindent

\end{abstract}


\maketitle

\section{Introduction}\label{intro}
The experimental investigation of ultracold atomic systems has made possible the realization of several celebrated models in quantum mechanics and condensed matter. These experiments are characterized by the rigorous control over the parameters of the system and by precise measurement techniques. The manipulation of optical traps, for instance, allows for the construction of different confining geometries \cite{bloch1}, from one-dimensional tubes \cite{weiss1,haller1} to lattice systems \cite{stoferle,greiner}. Traps consisting of two wells, in particular, have been extensively employed in recent experiments \cite{anderlini,folling,trotzky,greif,murmann}. They are of special interest in the study the Josephson effect \cite{josephson} in cold atoms, where Bose-Einstein condensates placed in such potentials \cite{ketterle_josephson,gerritsma} are considered in analogy to superconductors \cite{tinkham,anderson}. The high degree of precision demonstrated in these experiments also extends to the number of particles under consideration and to the strength of the interactions between them. Two-component fermionic systems with only a few atoms \cite{jochim1,jochim2,jochim3} and paradigmatic models such as the infinitely repulsive Tonks-Girardeau gas \cite{paredes,weiss2} can also be realized and studied in the lab. Combinig these features, other experiments have recently explored multicomponent strongly correlated gases \cite{fallani}, which have shown several exotic properties \cite{fallani_hall}. 

The limit of strong interactions has been a particularly favored starting point in the theoretical study of one-dimensional systems with internal degrees of freedom, due to the possibility of mapping the Hamiltonian to an effective spin chain \cite{deuretz1,artem2,pu1,xiaoling,deuretz3}. One of the key features of this mapping is that the exchange coefficients of the spin chain are solely determined by the trapping potential, and powerful numerical methods to calculate these coefficientes are now available \cite{deuretz_mdist,conan}. Approaching the problem of few atoms in a trap in the strongly interacting regime has also provided knowledge of the fundamental properties of quantum magnetism \cite{oelkers,guan2,deuretz2,artem1,amin2,massign}. On the other hand, the many-body case in the limit of total population imbalance $-$ the ``impurity" problem $-$ also presents many interesting features, such as quantum flutter \cite{mathy} and Bloch oscillations in the absence of a lattice \cite{bloch_gangardt,bloch_gangardt_2,bloch_gamayun,yang_imp,pu_onebody} (the latter having been recently observed in experiments \cite{meinert}).

In this work we study a strongly interacting system composed of an impurity and a bosonic background in single- and double-well potentials. Different methods have been employed to study the many-body problem of bosons in a double-well, even outside the mean-field regime \cite{anglin,tonel,links,masiello_mctdh,garcia_busch}. Addressing the subject from a few-body perspective \cite{zollner,brouzos_dw,sowinski1,sowinski2,tylutki}, however, might lead to new insight on the properties of these systems. Here we show that, in the regime where the repulsion between the impurity and the background is dominant, the system can exhibit non-trivial dynamical effects: the impurity undergoes Josephson-like oscillations when initialized at the edge of the system, and can have its tunneling enhanced when a barrier is present. These effects provide new perspectives in the study of spin state transfer and quantum transport in one-dimensional systems, and should be observable using current experimental techniques.

\section{System description and Hamiltonian}\label{description}

\begin{figure}[H]
\centering
\includegraphics[scale=0.5]{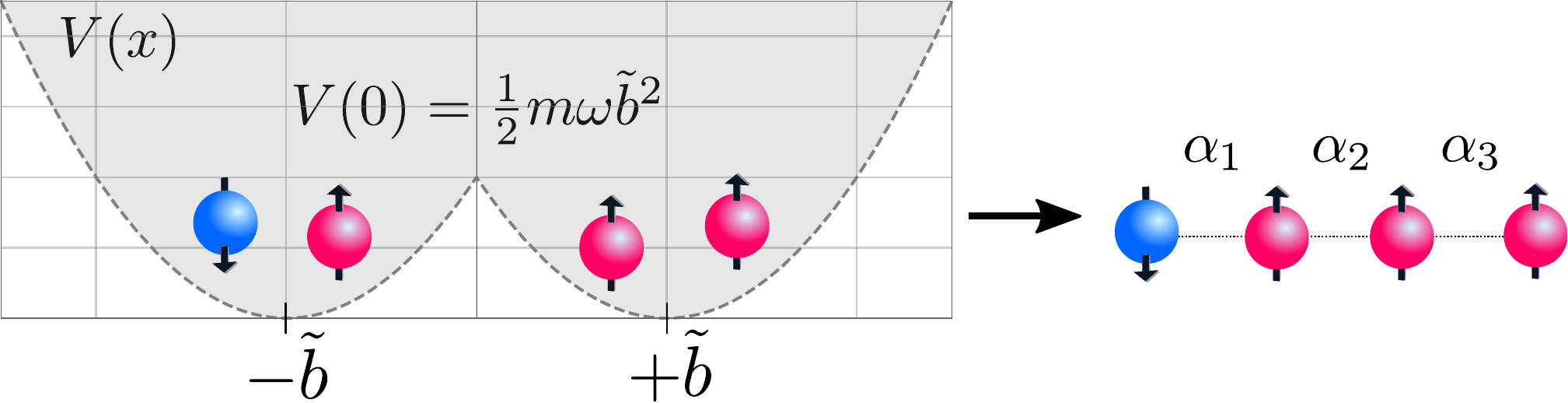}
\caption{(color online) Sketch of the 1D bosonic system with an impurity in the double-well potential. The parameter $\tilde{b}$ sets the position of the minimum of each well and also the size of the barrier between them at $x=0$. In the limit of strong interactions, the system can be mapped to a spin chain where the coefficients $\alpha$ are determined by the shape of the trap.}
\label{fig1}
\end{figure}

We consider the problem of an impurity confined in the presence of a background of strongly interacting bosons. We assume the impurity is a boson in an internal state defined by $\vert\downarrow\rangle$, while the remaining identical bosons are described by $\vert\uparrow\rangle$. Consequently, all atoms have the same mass $m$. Two-component Bose gases can be realized experimentally using, for instance, ${}^{87}$Rb atoms in different hyperfine states such as $\vert F=2,m_F=-1\rangle$ and $\vert F=1,m_F=1\rangle$ \cite{erhard,widera}. The number of identical bosons is given by $N_\uparrow$, while the total system size is $N=N_\uparrow + 1$. The Hamiltonian for this problem is written as
\begin{equation}\label{hm2}
H=\sum_{i=1}^{N} H_0(x_i)+g\sum_{i=1}^{N_\uparrow}\delta(x_\downarrow-x_{\uparrow i})+\kappa g \sum_{i<j}^{N_\uparrow} \delta(x_{\uparrow i}-x_{\uparrow j}),
\end{equation}
where the first sum involves the single-particle Hamiltonian $H_0$ (see below) which is the same for both components, while the remaining terms account for the contact interactions. The coordinates are denoted by $x_\downarrow$ for the impurity and $x_{\uparrow i}$ for the remaining bosons. The interaction strength is defined by $g$ for the impurity-background interactions, and as $\kappa g$ for the background-background interactions. Those parameters can be experimentally manipulated using Feshbach \cite{feshbach} or confinement induced resonances \cite{olshanii}. The single-particle Hamiltonian in Eq.~\ref{hm2} is given by $H_0(x)=-\frac{\hbar^2}{2m}\frac{\partial^2}{\partial x^2}+V(x)$. Here, $V(x)$ is a double-well potential (see Fig.~\ref{fig1}) expressed as $V(x)=\frac{1}{2}m\omega^2\left(\vert x \vert -\tilde{b} \right)^2$, where $\omega$ is the trapping frequency. The parameter $\tilde{b}$ denotes the displacement of the two minima of the wells with respect to the origin, and also defines the size of the barrier at this point as $V(0)=\frac{1}{2}m \omega^2 \tilde{b}^2$. We then refer to $\tilde{b}$ as the ``barrier parameter". By making $\tilde{b}=0$ we naturally recover the harmonic single-well potential. Although this form of potential has analytical solutions in terms of parabolic cylinder functions \cite{merzbacher}, we obtain the single-particle wave functions and energies through numerical diagonalization (see \ref{orbitals}). We will focus on the behavior of the spatial distributions and the impurity dynamics in the repulsive case ($g,\kappa>0$), for different choices of the intraspecies interaction parameter $\kappa$ and the barrier parameter $\tilde{b}$. While cases of attractive interactions ($g,\kappa<0$) can in principle be explored, the properties of the system in this regime reproduce only highly excited states related to the so-called Super Tonks-Girardeau gas \cite{ST,fermi_ST}. Simulating the dynamics of systems with attractive interactions would likely require taking into account the formation of bound pairs, an effect which is beyond the scope of the formalism employed here.
Throughout this work, we will consider all quantities in harmonic oscillator units; therefore, length, energy and time are given in units of $l=\sqrt{\hbar/m\omega}$, $\hbar \omega$ and $\omega^{-1}$, respectively. While the intraspecies interaction parameter $\kappa$ is dimensionless, the parameter $g$ is considered in units of $\hbar^2/m l$. For simplicity, we also make the barrier parameter dimensionless by rescaling it as $b=\tilde{b}/l$. In our calculations, we set $g=20$ and assume that $\hbar=\omega=m=1$.

In the limit of strong interactions ($g\gg 1$), Hamiltonian \ref{hm2} can be written, up to linear order in $1/g$, as the XXZ spin chain (see \ref{mapping} for details)
\begin{equation}\label{pert}
H_s=E_0 \mathbb{1}-\sum_{i=1}^{N-1}\frac{\alpha_i}{g}\left[\frac{1}{2}(\mathbb{1}-\vec{\sigma}^i\cdot\vec{\sigma}^{i+1})+\frac{1}{\kappa}(\mathbb{1}+\sigma_z^{i}\sigma_z^{i+1})\right],
\end{equation}
where $E_0$ is the energy of the system in the limit of infinite repulsion and $\sigma^{i}$ denotes a Pauli matrix acting on site $i$. The coefficients $\alpha_i$, often called geometric coefficients, are calculated using the wave function for a system of $N$ spinless fermions, which is constructed as the Slater determinant of the lowest occupied orbitals in the trap (see \ref{geometric}). The spatial part of the wave function for a bosonic system is then obtained by means of the Fermi-Bose mapping \cite{girardeau}. A comparative study of the spatial distributions for a strongly interacting few-body bosonic system in the double-well has been presented in \cite{garcia_polls}.

\section{Spin densities}
\begin{figure}[H]
\centering
\includegraphics[scale=0.45]{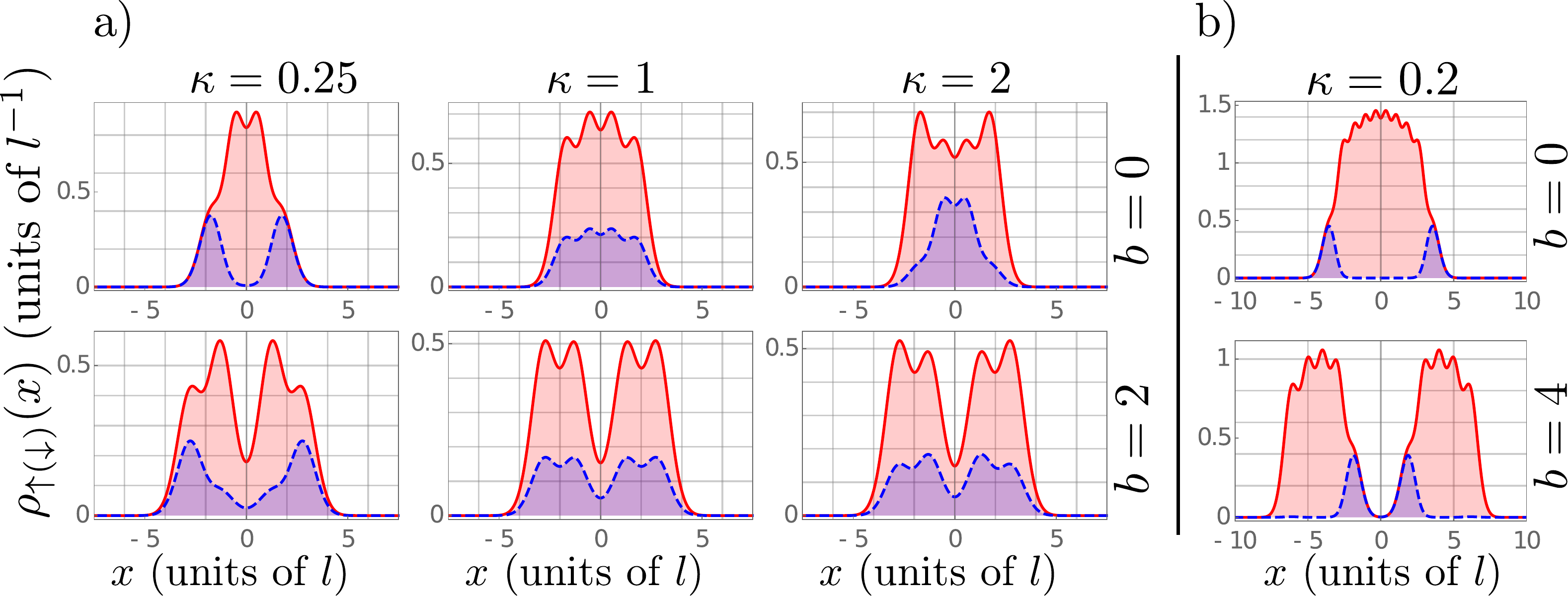}
\caption{(color online) a) Spin densities for the ground state of a 3+1 system at different values of the barrier parameter $b$ and of the background interaction parameter $\kappa$ (the values for each panel are determined by the labels on the rows/columns). The red curve corresponds to the background density $\rho_\uparrow (x)$ while the blue curve corresponds to the impurity density $\rho_\downarrow (x)$. b) Spin densities for a 9+1 system at $\kappa=0.2$ and two different choices of $b$. For a larger value of $b$, the impurity has a greater probability of being placed near the barrier, at the center of the system. The density for each component is normalized to its corresponding number of particles.}
\label{fig2}
\end{figure}

To obtain the probability densities for each component in the system, we must combine the spatial distributions of the atoms in the trap with the probability of magnetization of the corresponding site for an eigenstate of the spin chain described by Eq.~\ref{pert}. The {\it spin densities} are therefore given by 

\begin{equation}
\rho_{\uparrow(\downarrow)}(x)=\sum_{i=1}^{N}\rho^{i}_{\uparrow(\downarrow)}(x),
\end{equation}
with $\rho^i_{\uparrow(\downarrow)}=m^{i}_{\uparrow(\downarrow)}\rho^{i}(x)$, where $\rho^i(x)$ describes the individual atomic densities (see \ref{spatial}), while $m^{i}_{\uparrow(\downarrow)}$ denotes the probability of each site in a spin wave function $\vert \chi \rangle$ having spin up or down. The quantities $\rho_\downarrow(x)$ and $\rho_\uparrow(x)$ thus describe the spatial distributions of the impurity and the background bosons, respectively. In Fig.~\ref{fig2} a) we show the results for the spin densities of a 3+1 system for $b=0$ (single-well) and $b=2$ (double-well) at different values of the intraspecies parameter $\kappa$.

The cases where $b=0$ correspond to the results expected for the harmonic trap, which have been broadly covered for bosonic and fermionic systems in previous works \cite{hao,guan,deuretz2,lindgren,pu1}. For $\kappa<1$ the repulsion between the impurity and the background is larger than the background repulsion. This causes the impurity to be pushed to the edges of the system, which is an effect also found in the case of a weakly interacting background \cite{amin1}. In this regime, the system exhibits Ising-type ferromagnetic correlations, and is characterized by a nearly degenerate ground state \cite{massign}. At $\kappa=1$ all interactions are equal and the spin densities show the Heisenberg-type ferromagnetic profiles expected for isospin bosons \cite{guan2}. In this case both distributions display the same characteristic Tonks-Girardeau spatial densities, but scaled to the number of particles in each species. When $\kappa>1$, the repulsion between the background bosons dominates, and we observe predominantly antiferromagnetic correlations, where the impurity is placed near the center of the trap.
In the double-well potential, all densities are depleted in the center of the system, but this is not the only relevant effect. For the case of $\kappa<1$ the impurity has now a larger probability of being near the center of the trap (as compared to the single-well case), since the background density is strongly reduced in this region. It is even possible to expect a configuration (specially for a larger number of background bosons) where the impurity is completely localized near the barrier between the wells (see Fig.\ref{fig2} b)). This effect is directly related to the imbalance in the numerical values of the geometrical coefficients at the edges and near the center of the system as $b$ is increased. For the cases of $\kappa=1$ , again the impurity and the background densities have the same shape, aside from normalization. At $\kappa>1$, we observe a similar configuration, with a small bias of the impurity toward the center of the trap.

\section{Dynamics}\label{dynamics}

We now turn to the dynamics of the impurity after being initialized at the left edge of the system. The corresponding spin state is therefore given by $\vert \downarrow\uparrow\uparrow\uparrow\rangle$. Since this is not an eigenstate of the spin chain, we can expect the spin state to evolve in time governed by Eq.~\ref{pert}, and we denote it by $\vert \chi(t)\rangle$. A thorough study of spin state transfer in traps of different shapes has been done by Volosniev {\it et. al.} in \cite{artem2}. It has been shown \cite{nikolopoulos,christandl,loft1} that transfer is optimized by considering $\kappa=2$ (which turns Eq.~\ref{pert} into an XX Hamiltonian) with a set of exchange coefficients where $\alpha_j\propto\sqrt{j(N-j)}$. Here we focus on the tunneling times for the impurity between the wells (or between the left and right sides of the system in the case of a single well) when the background repulsion is smaller than the repulsion between the impurity and the background ($\kappa<1$). We point out that, since we do not consider any other external perturbations, like trap or interaction quenches, we can assume that the spatial distributions remain in the ground state. This also allows us to consider only the manifold of Eq.~\ref{pert} with lowest energy. To quantify the dynamics of the impurity, we calculate its average position as 
\begin{equation}
\langle x_\downarrow(t)\rangle=\int \rho_\downarrow(x,t) \,x \,dx,
\end{equation}
where $\rho_\downarrow(x,t)$ is the time dependent spin density calculated with the spin state $\vert \chi(t)\rangle$. When considering the regime of $\kappa<1$, we observe that the projections of the initial wave function on the two-lowest  eigenstates are dominant when compared to the case of higher excited states. This allows us to attempt a two-level description for the time evolution of the spin wave function; we thus write $\vert \psi(t)\rangle=c_g e^{-i\omega_R t}\vert g\rangle+c_e\vert e \rangle$, where $\vert g\rangle$ and $\vert e\rangle$ denote the two egeinstates of the spin chain with lower energy, and $c_g$ and $c_e$ are the projections of the initial wave function over these states. The frequency $\omega_J=E_e-E_g$ is given by the gap between the energies of the first excited state $E_e$ and the ground  state $E_g$.

\begin{figure}
\centering
\includegraphics[scale=0.5]{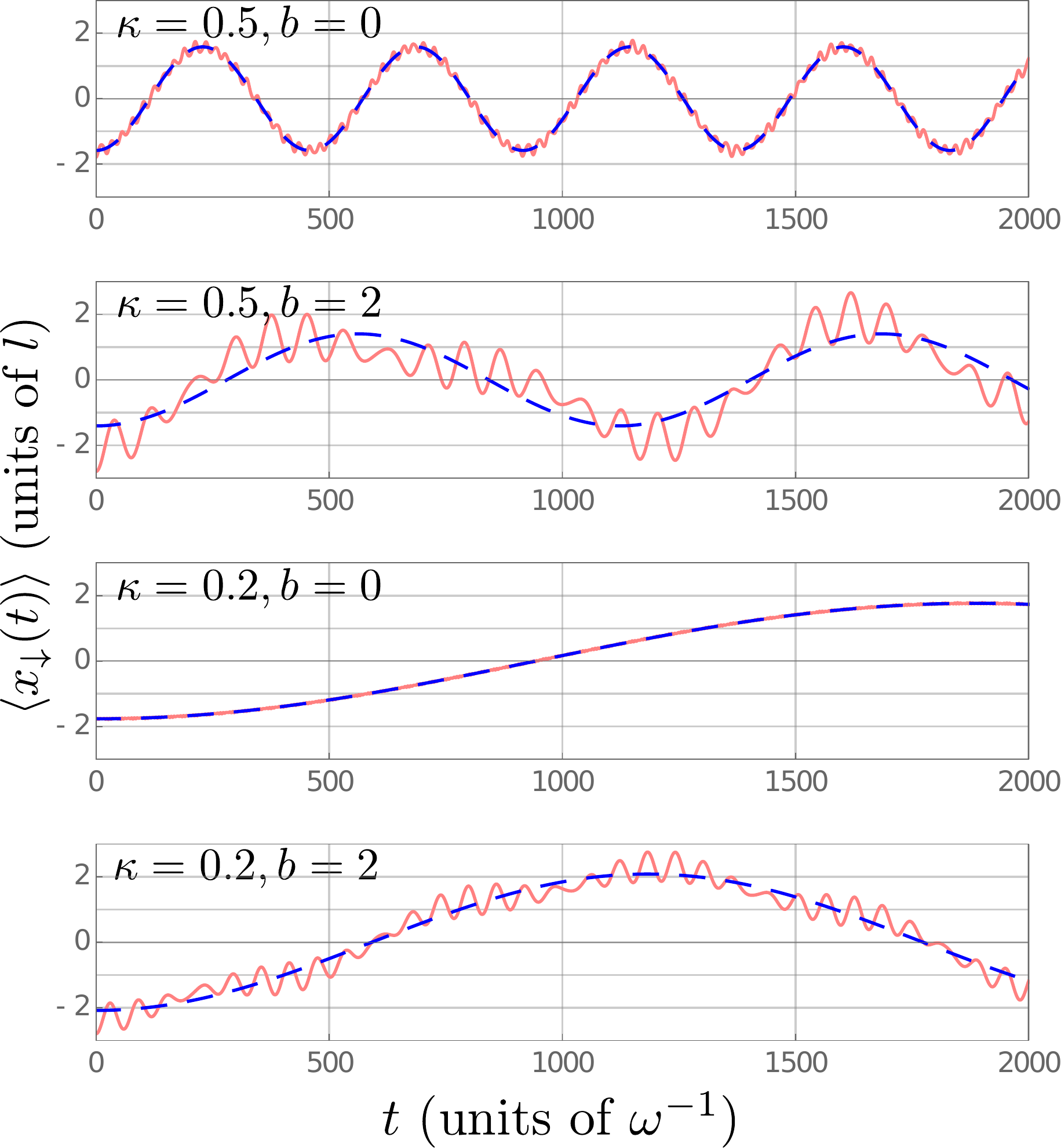}
\caption{(color online) Impurity average position as a function of time for different values of $\kappa$ and $b$. The solid red curves show the exact results, while the blue dashed curves are obtained with the two-level description (see text).}
\label{fig3}
\end{figure}

In Fig.~\ref{fig3}, we present the results for $\langle x_\downarrow(t)\rangle$ in the single-well ($b=0$) and double-well potentials ($b=2$), with two choices of $\kappa<1$, also showing a comparison with the two-level description in each of these cases. At $b=0$, we notice that the motion of the impurity between the edges of the system is very well captured by the two-level approximation. This behavior clearly resembles the oscillations in population expected for a bosonic Josephson junction described as a many-body system in a double-well. In the present context, however, the barrier is composed by the repulsive background gas. We expect these results to hold even in the case of more than one impurity, provided that the system is imbalanced (that is, the background gas must have a larger number of particles). In this situation, an initial state decribed by the minority species completely localized at either side of the trap should have its time evolution governed mainly by the two lowest energy states. In the single-well case with weaker intraspecies interaction ($\kappa=0.2$) the tunneling of the impurity is suppressed. Here, the the behavior of the background approaches that of an ideal Bose gas, where the atoms tend to ``bunch up" in the center of the trap.
Now, comparing the single- and double-well cases, we see that, for $\kappa=0.5$, the presence of the barrier slows down the tunneling of the impurity. Furthermore, we observe oscillations on a smaller scale, due to a larger overlap between the initial state and the excited states of the spin chain Hamiltonian. At $\kappa=0.2$, however, we get an enhanced tunneling of the impurity when considering a double-well as opposed to the single-well case. This effect has been also found with a different choice of double-well potential \cite{artem2}. One might interpret it as a splitting of the background gas by the barrier in such a way that the impurity is able to tunnel through faster than it would in the absence of the barrier. However, if we consider a single-particle problem where an atom is initialized in the left well, it is clear that increasing the barrier size would only lead to exponential suppression in the tunneling frequencies. We therefore conclude that the accelerated tunneling observed in the regimes we consider is only possible due to the presence of the bosonic background, and thus constitutes a many-body effect. 

\begin{figure}
\centering
\includegraphics[scale=0.5]{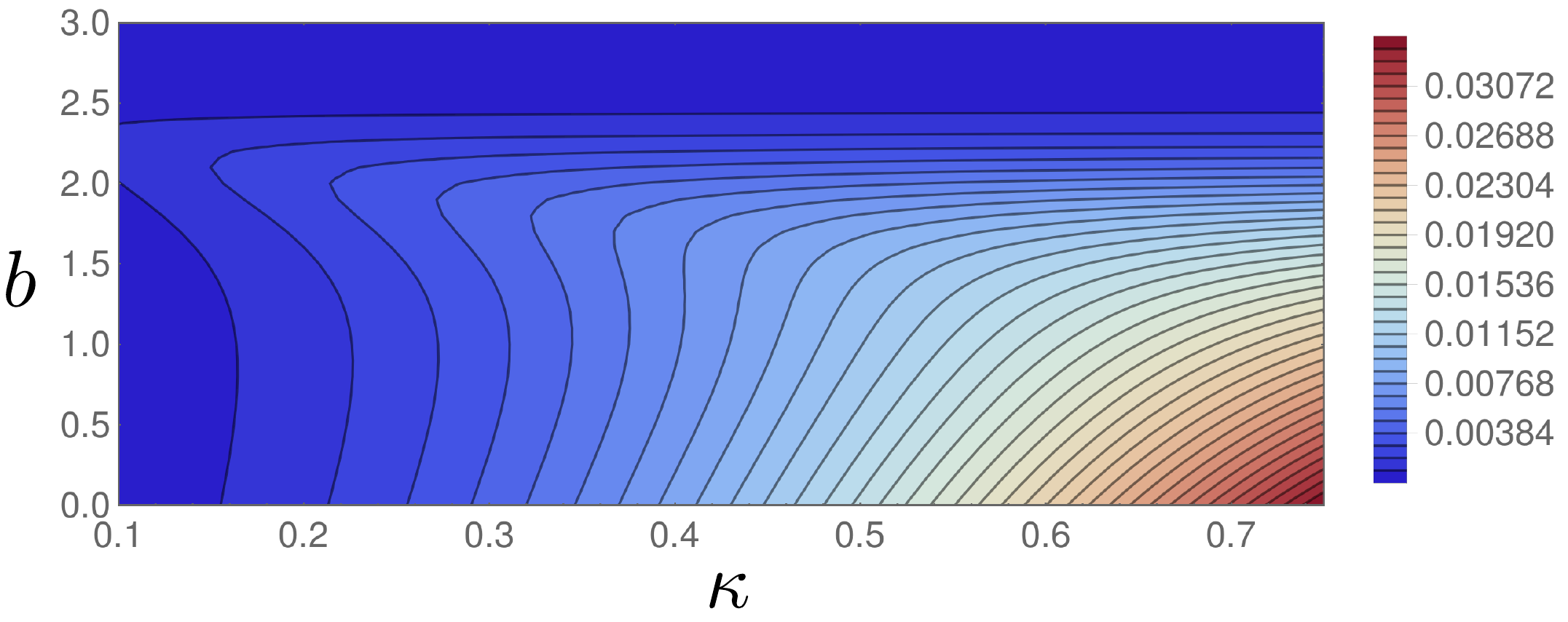}
\caption{(color online) Energy gap between the two lowest states of the spin chain for different values of the barrier parameter $b$ and background interaction $\kappa$ for the 3+1 case.}
\label{fig4}
\end{figure}

To get an understanding of this behavior over a larger parameter space, we plot in Fig.~\ref{fig4} the energy gap between the ground state and the first excited state for several values of $\kappa$ and $b$. The non-monotonic behavior of the gap as a function of $b$ indicates that, for small $\kappa$, there is some choice of barrier size that increases that energy gap, and therefore enables a higher tunneling frequency between the wells. As $\kappa$ increases, however, we see that this behavior disappears and the presence of the barrier only reduces the gap, thus making the tunneling slower.

\section{Increasing $N_\uparrow$}

\begin{figure}[H]
\centering
\includegraphics[scale=0.5]{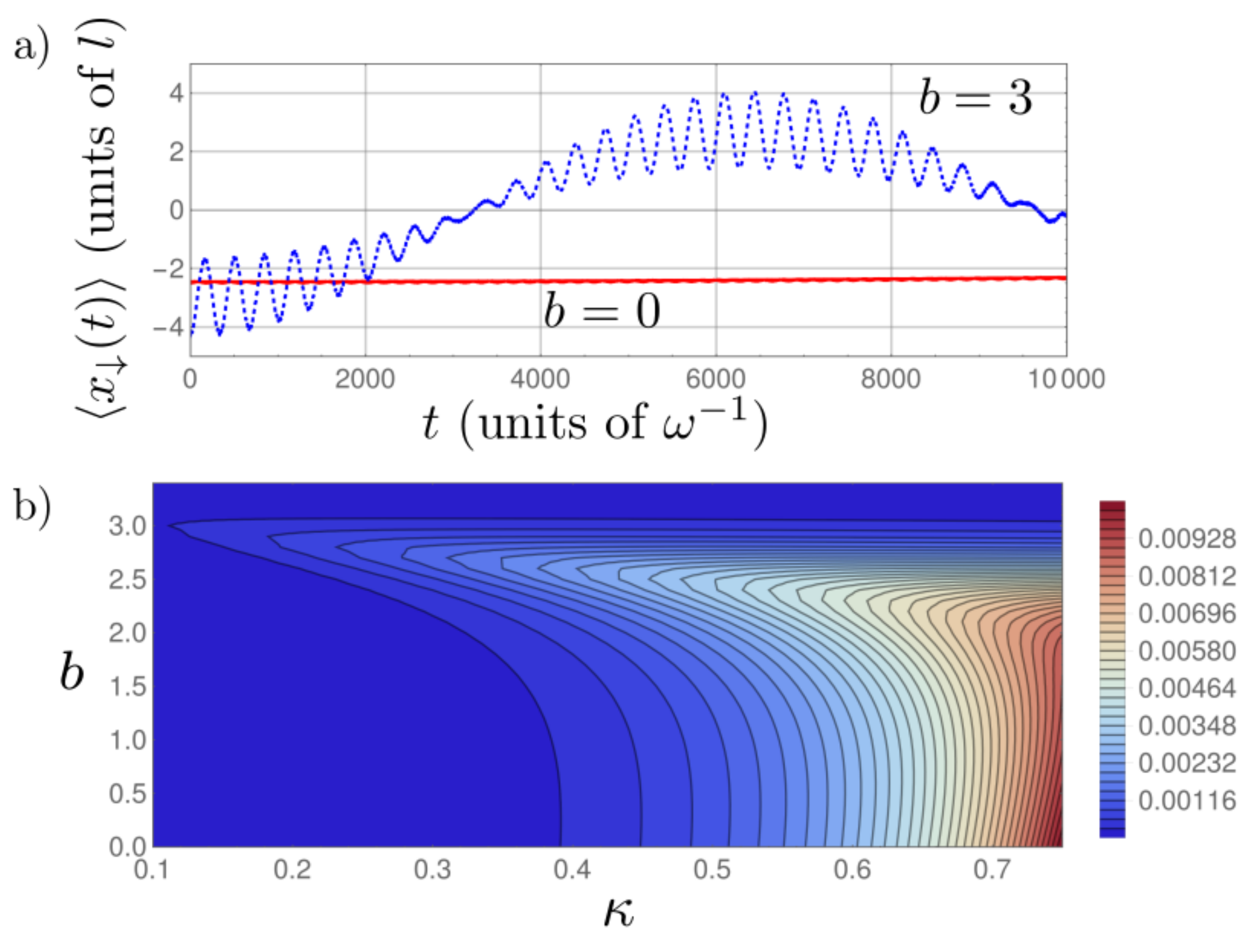}
\caption{(color online) a) Average position of the impurity as a function of time for the 5+1 case, in the single- ($b=0$, red solid curve) and double-well ($b=3$, blue dashed curve). The background interaction parameter is set as $\kappa=0.25$. b) Energy gap between the two lowest states of the spin chain for different values of the barrier parameter $b$ and background interaction $\kappa$.}
\label{fig5}
\end{figure}

As a final example, we consider a case where we increase the number of background bosons to $N_\uparrow=5$ (to observe similar effects as in the case of $N_\uparrow=3$, we choose to maintain an even total number of atoms). The initial state is once again defined by the impurity placed at the left edge of the system, that is, $\vert \downarrow\uparrow\uparrow\uparrow\uparrow\uparrow\rangle$. We keep the intraspecies repulsion parameter fixed at $\kappa=0.25$. In Fig.~\ref{fig5} a) we once again show the results for the average position of the impurity as a function of time. For the single-well, the tunneling times are so long that the impurity is effectively frozen at the left edge of the system. In this case, an analogy can be drawn to the self-trapped regime in a few-body system as presented in \cite{zollner}. For $b=3$, however, we again notice that a faster motion of the impurity from the left to the right well is induced. The difference in the results with and without the barrier is even clearer than in the case of $N_\uparrow=3$. This can also be seen in the energy gap between the two lowest states, as presented in Fig. \ref{fig5} b): a very pronounced curve shows the increase in this quantity for small $\kappa$ and $b>1$. We point out that the final time ($t=10^4$ in units of $\omega^{-1}$) considered in the present case is five times larger than in the case of $N_\uparrow=3$. Time scales for harmonically trapped systems are set by the inverse frequency, which in present experiments with few-body cold atoms is of around $100\mu$s \cite{jochim3}. The total times obtained in experimental setups can therefore be decreased by considering tighter traps.

\section{Conclusions}\label{conc}
We have studied the static and dynamic properties of an impurity in the presence of a background of bosons in single-well and double-well geometries. The ground state spin densities are described by a combination of the spatial distributions in the limit of infinite repulsion and the eigenstates of a spin chain. We have shown that the dynamics of an impurity initialized at the left edge of the system displays oscillations similar to the ones observed in Josephson junctions. Additionally, for weaker background interactions, the tunneling of the impurity can in fact be enhanced by the introduction of a barrier. This many-body effect is only possible in the presence of a background. We interpret it as the increase of the gap between the two lowest energy states, which governs the low-frequency dynamics of the system. Our results open new perspectives on the study of quantum transport in one-dimensional systems, hinting at the possibility of realizing a bosonic Josephson junction in the complete absence of an artificial barrier. Moreover, the inclusion and manipulation of double-well potentials and even lattices may allow for the optimized transfer of spin states.

\ack
The authors thank Xiaoling Cui and Artem Volosniev for reading and commenting on the manuscript, and Tomasz Sowiński for fruitful discussions. The following agencies - Conselho Nacional de Desenvolvimento Científico e Tecnológico (CNPq), the Danish Council for Independent Research DFF Natural Sciences and the DFF Sapere Aude program - are gratefully acknowledged for financial support. 

\appendix

\section{Exact solutions, geometrical coefficients and spatial distributions in the limit of infinite repulsion}

In this Appendix we present the solutions obtained with numerical diagonalization for the single-particle in a double-well, mapping between a strongly interacting one-dimensional system and the spin chain Hamiltonian described by Eq.~\ref{pert}, expressions for the geometrical coefficients and densities in the impenetrable limit, and the analytical form of the eigenstates and eigenvalues of the spin chain in the 3+1 case. The single-particle energies and numerical values for the exchange coefficients are calculated using the open-source code CONAN \cite{conan}. Some of these quantities depend on the spinless fermion wave function $\Phi(x_1,...,x_N)$, which is constructed as the Slater determinant of the $N$ lowest-lying orbitals of the trapping potential. The energy $E_0$ from Eq.~\ref{pert} is then simply the sum of the energies of these individual states.

\subsection{Single-particle solutions in the double-well}\label{orbitals}
\begin{figure}[H]
\centering
\includegraphics[scale=0.4]{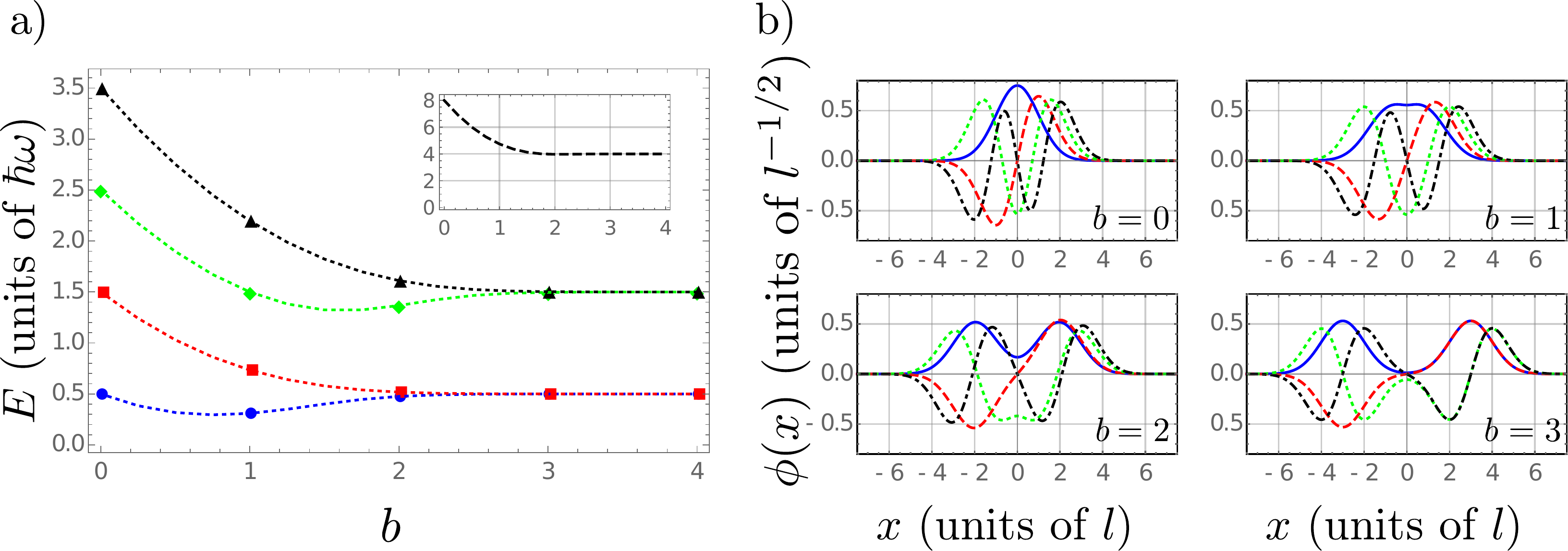}
\caption{a) Single-particle energies of the 4 lowest eigenstates of the double-well potential as a function of the barrier parameter $b$. The dotted curves show the results calculated with CONAN (see text), while the symbols are the values obtained by numerical diagonalization. The inset shows the energy $E_0$ of the spinless fermion wave function as a function of $b$ for $N=4$. b) The four lowest single-particle eigenstates for different values of the barrier parameter $b$. Solid blue, dashed red, dotted green and dot-dashed black curves correspond to ground state, 1st, 2nd and 3rd excited states, respectively.}
\label{figap1}
\end{figure}

The eigenvalues and eigenstates of a particle in a double-well were obtained through numerical diagonalization of the hamiltonian $H_0$ using the 50 lowest states of the harmonic oscillator trap ($b=0$) as basis. In Fig.~\ref{figap1} we present these solutions for different values of the barrier parameter $b$. In panel a), we show how each pair of states becomes degenerate as the barrier size is increased. This is reflected in the eigenstates shown in panel b): at larger values of $b$, the ground state and the first excited state have the same probability distribution, differing only in parity.

\subsection{Mapping the strongly interacting system to a spin chain Hamiltonian}\label{mapping}

In this section we show details of the mapping between Hamiltonian \ref{hm2} and the spin chain described by Eq.\ref{pert}. Although different approaches have been used to describe this mapping \cite{deuretz1,deuretz2,artem1,xiaoling}, we focus on the one presented in \cite{artem2}. We start by considering a more general bosonic Hamiltonian of the form
\begin{eqnarray}
H&=&\sum_{i=1}^{N} H_0(x_i)+g\sum_{i=1}^{N_\uparrow}\sum_{j=1}^{N_\downarrow}\delta(x_{\uparrow i}-x_{\downarrow j})+\kappa g \sum_{i<i'}^{N_\uparrow} \delta(x_{\uparrow i}-x_{\uparrow i'})\nonumber \\&+&\kappa g \sum_{j<j'}^{N_\downarrow} \delta(x_{\downarrow j}-x_{\downarrow j'}),
\end{eqnarray}
where the total number of particles is given by $N=N_\uparrow+N_\downarrow$. In the limit of infinite interactions ($g\rightarrow \infty$), the eigenstates of Hamiltonian \ref{hm2} can be described by
\begin{equation}\label{ansatz}
\Psi=\sum_{k=1}^{L(N_\uparrow,N_\downarrow)}a_k P_k\Phi_0(\{x_{\uparrow i},x_{\downarrow j}\}),
\end{equation} 
where the sum runs over the $L(N_\uparrow,N_\downarrow)=\binom{N_\uparrow+N_\downarrow}{N_\uparrow}$ permutations of the coordinates, and $P_k$ is the permutation operator. In this expression, $\Phi_0$ is simply the wave function in the impenetrable limit, with coordinates fixed as $\{x_{\uparrow i}\}$ and $\{x_{\downarrow j}\}$, with $i=1,...,N_\uparrow$ and $j=1,...,N_\downarrow$. To investigate the behavior of the energy at very strong (but finite) interactions, we use the Hellmann-Feynman theorem, which gives
\begin{eqnarray}\label{hellmann}
\frac{\partial E}{\partial g}&=&\sum_{i=1}^{N_\uparrow}\sum_{j=1}^{N_\downarrow}\langle\Psi\vert\delta (x_{\uparrow i}-x_{\downarrow j})\vert\Psi\rangle+\kappa \sum_{i<i'}^{N_\uparrow}\langle\Psi\vert\delta (x_{\uparrow i}-x_{\uparrow i'})\vert\Psi\rangle \nonumber \\&+&\kappa \sum_{j<j'}^{N_\downarrow}\langle\Psi\vert\delta (x_{\downarrow j}-x_{\downarrow j'})\vert\Psi\rangle,
\end{eqnarray}
where the first term on the right-hand side accounts for interactions between different bosons, while the remaining terms arise from interactions between identical bosons. The conditions for the derivatives at the contact point between two particles are given by
\begin{eqnarray}\label{cond1}
\left(\frac{\partial\Psi}{\partial x_{\uparrow i}}-\frac{\partial\Psi}{\partial x_{\uparrow i'}}\right)\Bigg\rvert_-^+=2\kappa g \Psi(x_{\uparrow i}=x_{\uparrow i'}),\nonumber \\
\left(\frac{\partial\Psi}{\partial x_{\downarrow j}}-\frac{\partial\Psi}{\partial x_{\downarrow j'}}\right)\Bigg\rvert_-^+=2\kappa g \Psi(x_{\downarrow j}=x_{\downarrow j'})
\end{eqnarray}
for identical bosons and 
\begin{equation}\label{cond2}
\left(\frac{\partial\Psi}{\partial x_{\uparrow i}}-\frac{\partial\Psi}{\partial x_{\downarrow j}}\right)\Bigg\rvert_-^+=2g \Psi(x_{\uparrow i}=x_{\downarrow j}),
\end{equation}
for a distinguishable pair. In the expressions above we have $+\rightarrow x_m-x_n= 0^+$, while $-\rightarrow x_m-x_n=0^-$.

Combining Eqs.\ref{hellmann}, \ref{cond1} and \ref{cond2}, we get
\begin{equation}
\frac{\partial E}{\partial g}=\frac{K_{\uparrow\downarrow}}{g^2}+\frac{K_{\uparrow\uparrow}}{\kappa g^2}+\frac{K_{\downarrow\downarrow}}{\kappa g^2},
\end{equation} 
where
\begin{eqnarray}
K_{\uparrow\downarrow}=\frac{\sum_{i=1,j=1}^{N_\uparrow,N_\downarrow}\int dx_{\uparrow 1},\cdots,dx_{\uparrow N_\uparrow}\int dx_{\downarrow 1},\cdots,dx_{\downarrow N_\downarrow}\Bigg\rvert\left(\frac{\partial\Psi}{\partial x_{\uparrow i}}-\frac{\partial\Psi}{\partial x_{\downarrow j}}\right)\bigg\rvert_-^+\Bigg\rvert^2\delta (x_{\uparrow i}-x_{\downarrow j})}{4\int dx_{\uparrow 1},\cdots,dx_{\uparrow N_\uparrow}\int dx_{\downarrow 1},\cdots,dx_{\downarrow N_\downarrow}\vert \Psi \vert^2}\nonumber,\\
K_{\uparrow\uparrow}=\frac{\sum_{i<i'}^{N_\uparrow}\int dx_{\uparrow 1},\cdots,dx_{\uparrow N_\uparrow}\int dx_{\downarrow 1},\cdots,dx_{\downarrow N_\downarrow}\Bigg\rvert\left(\frac{\partial\Psi}{\partial x_{\uparrow i}}-\frac{\partial\Psi}{\partial x_{\uparrow i'}}\right)\bigg\rvert_-^+\Bigg\rvert^2\delta (x_{\uparrow i}-x_{\uparrow i'})}{4\int dx_{\uparrow 1},\cdots,dx_{\uparrow N_\uparrow}\int dx_{\downarrow 1},\cdots,dx_{\downarrow N_\downarrow}\vert \Psi \vert^2}\nonumber,\\
K_{\downarrow\downarrow}=\frac{\sum_{j<j'}^{N_\downarrow}\int dx_{\uparrow 1},\cdots,dx_{\uparrow N_\uparrow}\int dx_{\downarrow 1},\cdots,dx_{\downarrow N_\downarrow}\Bigg\rvert\left(\frac{\partial\Psi}{\partial x_{\downarrow j}}-\frac{\partial\Psi}{\partial x_{\downarrow j'}}\right)\bigg\rvert_-^+\Bigg\rvert^2\delta (x_{\downarrow j}-x_{\downarrow j'})}{4\int dx_{\uparrow 1},\cdots,dx_{\uparrow N_\uparrow}\int dx_{\downarrow 1},\cdots,dx_{\downarrow N_\downarrow}\vert \Psi \vert^2}\nonumber ,
\end{eqnarray}
where the denominator introduces a normalization factor.
Integrating with respect to $g$ we obtain the following energy functional 
\begin{equation}
E=E_0-\left(\frac{K_{\uparrow\downarrow}}{g}+\frac{K_{\uparrow\uparrow}}{\kappa g}+\frac{K_{\downarrow\downarrow}}{\kappa g}\right),
\end{equation} 
where $E_0$ is the energy in the limit of infinite repulsion, and we neglect terms of higher order in ($1/g$). By introducing the wave function described by Eq.\ref{ansatz} in the expression above, we obtain
\begin{eqnarray}\label{functional}
E=E_0-\nonumber \\ 
\frac{\sum_{i=1}^{N-1}\frac{\alpha_i}{g}\left(\sum_{k=1}^{L(N_\uparrow-1,N_\downarrow-1)}A^{\uparrow\downarrow}_{ik}+\frac{2}{\kappa}\sum_{k=1}^{L(N_\uparrow-2,N_\downarrow)}A_{ik}^{\uparrow\uparrow}+\frac{2}{\kappa}\sum_{k=1}^{L(N_\uparrow,N_\downarrow-2)}A_{ik}^{\downarrow\downarrow}\right)}{\sum_{k=1}^{L(N_\uparrow,N_\downarrow)}a_{k}^2}\nonumber\\
\end{eqnarray}
with
\begin{eqnarray}\label{acoef}
A_{ik}^{\uparrow\downarrow}=(a_{ik}^{\uparrow\downarrow}-a_{ik}^{\downarrow\uparrow})^2,\,\,\,\,\,\,\,\,\,
A_{ik}^{\uparrow\uparrow}=(a_{ik}^{\uparrow\uparrow})^2,\,\,\,\,\,\,\,\,\,
A_{ik}^{\downarrow\downarrow}=(a_{ik}^{\downarrow\downarrow})^2,
\end{eqnarray}
where $a_{ik}^{\uparrow\downarrow}$ represents the coefficients in Eq. \ref{ansatz} multiplying terms with nighboring $\uparrow$ and $\downarrow$ particles at position $i$ and $i+1$, while the remainig terms have the same role, for $\downarrow\uparrow$, $\uparrow\uparrow$ and $\downarrow\downarrow$ pairs. The purpose of such terms is to account for the energy contribution of exchanging two neighboring particles with particular spin projections. The coefficients $\alpha_i$ are now independet of spin, and can be written as
\begin{equation}\label{geo}
\alpha_i=\frac{\int_{x_1<x_2\cdots<x_N-1}dx_1...dx_{N-1}\Big|\frac{\partial \Phi_0(x_1,\cdots,x_i,\cdots,x_N)}{\partial x_N}\Big|^2_{x_N=x_i}}{\int_{x_1<x_2\cdots<x_N-1}dx_1\cdots dx_N |\Phi_0(x_1,\cdots,x_i,\cdots,x_N)|^2},
\end{equation}
where $\Phi_0(x_1,\cdots,x_i,\cdots,x_N)$ is again the wave function present in Eq.~\ref{ansatz} (where we have ommited the spin indices). Since this wave function is defined in the region determined by a particular order of the coordinates, it is enough to calculate the integrals in one particular sector, such as $x_1<x_2\cdots<x_N-1$.

Now, let us consider a spin chain Hamiltonian defined as
\begin{equation}\label{spinh}
H_s=E_0-\sum_{i=1}^{N-1}J_i \left(\Pi_{\uparrow\downarrow}^{i,i+1}+\frac{1}{\kappa}\Pi_{\uparrow\uparrow}^{i,i+1}+\frac{1}{\kappa}\Pi_{\downarrow\downarrow}^{i,i+1}\right)
\end{equation}
where $\Pi_{\uparrow\downarrow}^{i,i+1}=\frac{1}{2}(\mathbb{1}-\vec{\sigma}^i\cdot\vec{\sigma}^{i+1})$ is the operator that exchanges neighboring spins with different projections and $\Pi_{\uparrow\uparrow}^{i,i+1}=\Pi_{\downarrow\downarrow}^{i,i+1}=\frac{1}{2}(\mathbb{1}+\sigma_z^{i}\sigma_z^{i+1})$ have the same action, but for identical spins. A generic spin state can now be written as
\begin{equation}
\vert \chi \rangle=\sum_{k=1}^{L(N_\uparrow,N_\downarrow)}a_k P_k \vert \uparrow_1\cdots \uparrow_{N_\uparrow}\downarrow_1\cdots\downarrow_{N_\downarrow}\rangle,
\end{equation}
where once again the sum runs over the permutations of the $N_\uparrow$ and $N_\downarrow$ spins. If we calculate the expected value of Hamiltonian~\ref{spinh} as $\langle \chi \vert H\vert \chi\rangle$, we obtain
\begin{eqnarray}\label{functional2}
\langle \chi \vert H\vert \chi\rangle=E_0-\nonumber \\ 
\frac{\sum_{i=1}^{N-1}J_i\left(\sum_{k=1}^{L(N_\uparrow-1,N_\downarrow-1)}A^{\uparrow\downarrow}_{ik}+\frac{2}{\kappa}\sum_{k=1}^{L(N_\uparrow-2,N_\downarrow)}A_{ik}^{\uparrow\uparrow}+\frac{2}{\kappa}\sum_{k=1}^{L(N_\uparrow,N_\downarrow-2)}A_{ik}^{\downarrow\downarrow}\right)}{\sum_{k=1}^{L(N_\uparrow,N_\downarrow)}a_{k}^2}\nonumber \\
\end{eqnarray}
where the coefficients $A_{ik}^{\uparrow\downarrow}$, $A_{ik}^{\uparrow\uparrow}$ and $A_{ik}^{\downarrow\downarrow}$ have the same meaning as in Eq.~\ref{acoef}, and $\sum_{k=1}^{L(N_\uparrow,N_\downarrow)}a_{k}^2$ introduces a normalization factor. It becomes clear that the energy functionals given by Eqs. \ref{functional} and \ref{functional2} are identical if $J_i=\alpha_i/g$. Furthermore, by rewriting Eq.~\ref{spinh} in terms of the Pauli matrices, we obtain 
\begin{equation}
H_s=E_0\mathbb{1}-\sum_{i=1}^{N-1}\frac{\alpha_i}{g} \left[\frac{1}{2}(\mathbb{1}-\vec{\sigma}^i\cdot\vec{\sigma}^{i+1})+\frac{1}{\kappa}(\mathbb{1}+\sigma_z^{i}\sigma_z^{i+1})\right],
\end{equation}
which is the spin chain Hamiltonian described in Eq. \ref{pert} of the main text. We conclude then that the eigenvalue problems defined with Eqs.~\ref{functional} and \ref{functional2} are identical, which validates, for a strongly interacting system, the mapping between Hamiltonians \ref{hm2} and \ref{pert}.

\subsection{Geometrical coefficients}\label{geometric}

\begin{figure}[H]
\centering
\includegraphics[scale=0.5]{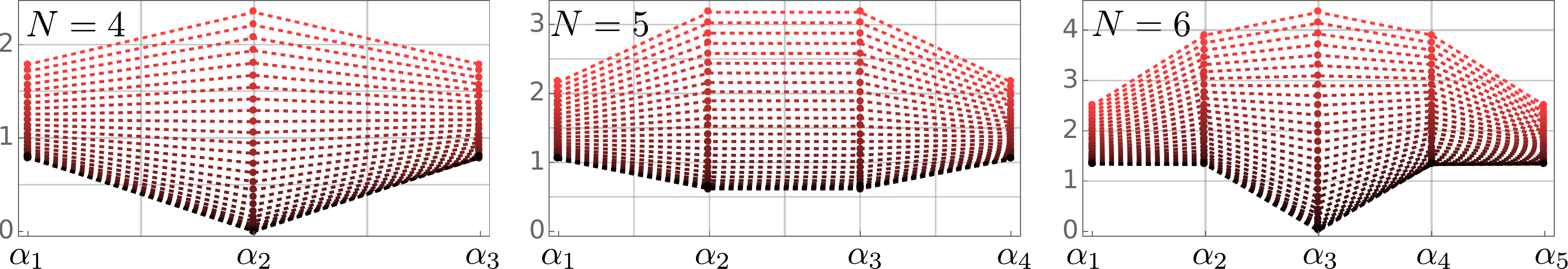}
\caption{Geometrical coefficients for three different particle numbers in a double-well. The circles mark the numerical values of each coefficient. The dotted lines connect the coefficients corresponding to the same values of $b$, which is increased from $b=0$ (top - lighter colors) to $b=3$ (bottom - darker colors).}
\label{figap2}
\end{figure}

Different methods have been developed for calculating the coefficients in Eq.~\ref{geo}, in particular exploring the determinant form of $\Phi_0(x_1,...,x_N)$. Here, we use the open-source code CONAN \cite{conan} to calculate them, considering the double-well potential with different barrier parameters. An equivalent approach, based on Chebyshev polynomials, has been published in \cite{deuretz_mdist}. In Fig.~\ref{figap2}, we show results for the geometrical coefficients in the cases of $N=4,5$ and 6. Due to the parity invariance of the trapping potentials considered here, we have $\alpha_N=\alpha_1,\alpha_N-1=\alpha_2,...$, which means we must calculate, at most, $N/2$ coefficients in each case.

In the single-well cases ($b=0$), the coefficients are simply the ones expected for the given number of particles in a harmonic trap. Particularly, for $N=4$ we have $\alpha_1\approx 1.78,\alpha_2\approx 2.34$. As $b$ is increased, the major change is observed for even $N$ in the central coefficient, which vanishes for $b\geqslant3$. In this limit, we have the two sides of the system described by almost completely separate harmonic traps with $N/2$ in each well. The values of the coefficients for harmonic traps in this situation have analytical expressions, given by $\alpha={{\pi}/{2}}$ for $N=2$ and $\alpha={3^3}/({2^3 \sqrt{2\pi}})$ for $N=3$. The situation is not the same for odd $N$. In this case, there is no central coefficient to vanish as the barrier is increased. We would expect such a system to exhibit different spatial densities and dynamics.

\subsection{Spatial correlations in the impenetrable limit}\label{spatial}

\begin{figure}
\centering
\includegraphics[scale=0.5]{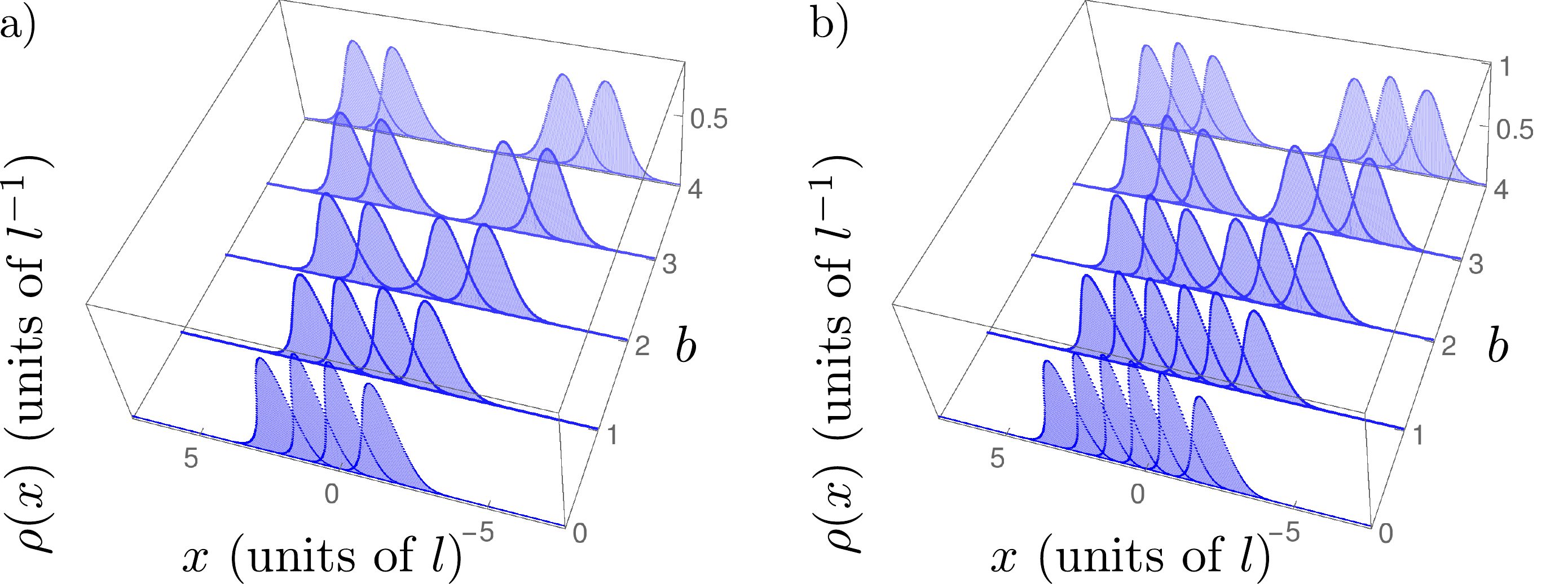}
\caption{Spatial distributions in the impenetrable limit for the case of a) $N=4$ and b) $N=6$.}
\label{figap3}
\end{figure}

In this section we describe the spatial densities for a given number of atoms $N$ in the impenetrable limit. These densities reproduce the results expected for a Tonks-Girardeau gas or a gas of spinless fermions, and are characterized by a chain of localized atoms. The individual distributions are calculated with the following expression \cite{deuretz2}
\begin{equation}\label{onebody}
\rho^i(x)=\int_{\Gamma} dx_1...dx_N \,\delta(x_i-x)|\Phi_0(x_1,...,x_i,...,x_N)|^2,
\end{equation} 
where the integral is performed in the region $\Gamma=x_1 <x_2<...<x_N$. The quantity $\rho^i(x)$ then gives the spatial distribution of the atom with index $i$. A formula for calculating these densities has been obtained by Deuretzbacher {\it et. al.} \cite{deuretz1}, and is written as
\begin{equation}\label{dets}
\rho^i(x)=\frac{\partial}{\partial x}\left( \sum_{j=0}^{N-1} c_{j}^{i} \frac{\partial^j}{\partial \lambda^j}\det \left[B(x)-1\lambda \right]\vert_{\lambda=0}\right),
\end{equation}
where $c_{j}^{i}=\frac{(-1)^{N-1}(N-j-1)!}{(i-1)!(N-j-i)!j!}$ and the matrix $B(x)$ is composed by the single-particle states superpositions $b_{mn}(x)=\int_{-\infty}^{x}dy\,\varphi_m(y)\varphi_n(y)$. In Fig.~\ref{figap3} we show these spatial distribution for the cases of $N=4$ and $N=6$ as the barrier parameter $b$ is increased.

\subsection{Eigenvalues and eigenstates of the spin chain Hamiltonian}\label{eigen}
We present here the analytical expression for the matrix form of Hamiltonian $H_s$ for the case of $N_\uparrow=3+N_\downarrow=1$. The choice of basis is given by $\{\vert\uparrow\uparrow\uparrow\downarrow\rangle,\vert\uparrow\uparrow\downarrow\uparrow\rangle,\vert\uparrow\downarrow\uparrow\uparrow\rangle,\vert\downarrow\uparrow\uparrow\uparrow\rangle\}$:
\begin{eqnarray}
H_s=\left(
\begin{array}{cccc}
 \frac{\kappa  \alpha _1+2 \left(\alpha _1+\alpha _2\right)}{g \kappa } & -\frac{\alpha _1}{g} & 0 & 0 \\
 -\frac{\alpha _1}{g} & \frac{2 \alpha _1+\kappa  \left(\alpha _1+\alpha _2\right)}{g \kappa } & -\frac{\alpha _2}{g} & 0 \\
 0 & -\frac{\alpha _2}{g} & \frac{2 \alpha _1+\kappa  \left(\alpha _1+\alpha _2\right)}{g \kappa } & -\frac{\alpha _1}{g} \\
 0 & 0 & -\frac{\alpha _1}{g} & \frac{\kappa  \alpha _1+2 \left(\alpha _1+\alpha _2\right)}{g \kappa } \\
\end{array}
\right)
\end{eqnarray}

\begin{figure}[H]
\centering
\includegraphics[scale=0.7]{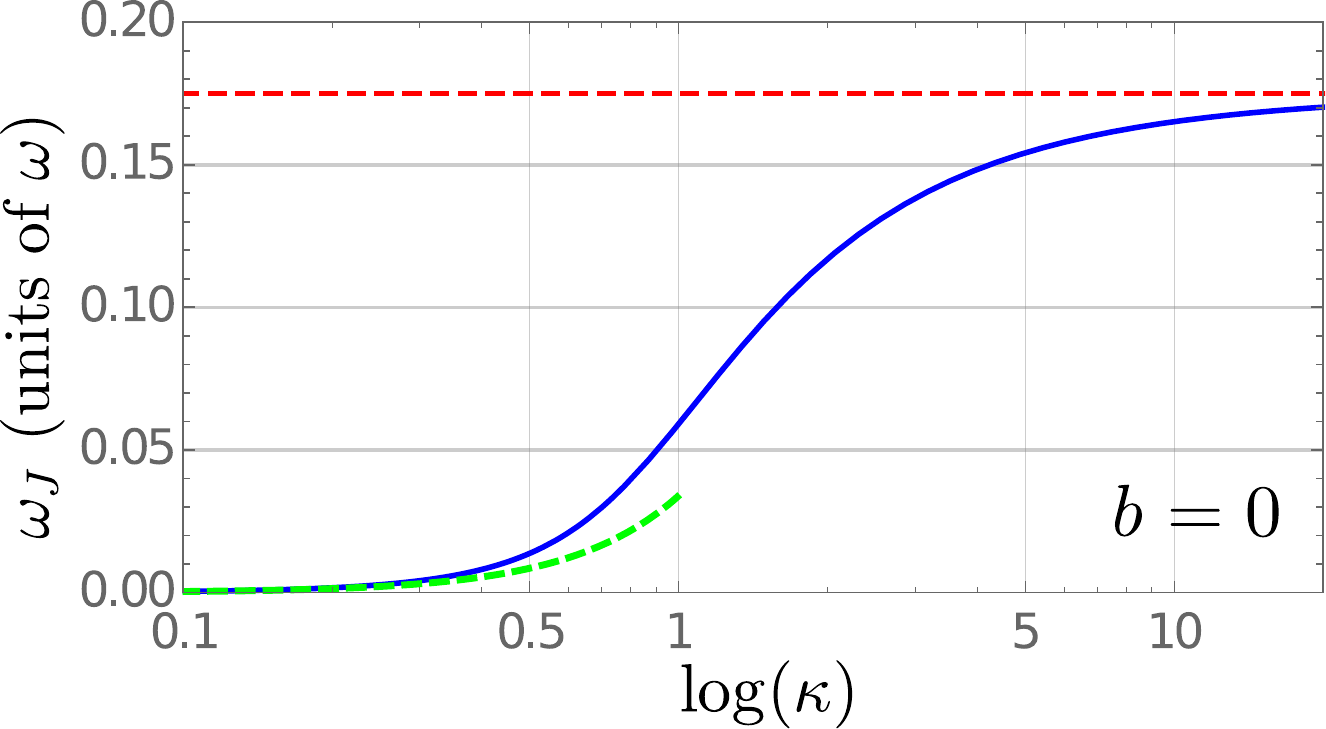}
\caption{Frequency $\omega_J$ as a function of the intraspecies parameter $\kappa$ in the single-well potential ($b=0$), for the case 3+1 (blue solid line). The dashed curves show the behavior of the frequency for small (green) and large (red) $\kappa$ as given by the analytical expressions in the text. In these results, we have used $\alpha_1=\alpha_3\sim 1.78$, $\alpha_2\sim2.34$ and $g=20$.}
\label{figap4}
\end{figure}

In the expression above we made the simplification that, due to the parity invariance of the trapping potential, $\alpha_3=\alpha_1$. Diagonalizing this matrix we obtain the following eigenvalues
\begin{eqnarray}
\epsilon_1&=& \frac{-\sqrt{\alpha _1^2 \kappa ^2+\alpha _2^2}+\alpha _1 (\kappa +2)+\alpha _2}{g \kappa },\\
\epsilon_2&=& \frac{\sqrt{\alpha _1^2 \kappa ^2+\alpha _2^2}+\alpha _1 (\kappa +2)+\alpha _2}{g \kappa },\nonumber\\
\epsilon_3&=& \frac{-\sqrt{\alpha _1^2 \kappa ^2+\alpha _2^2 (\kappa -1)^2}+\alpha _1 (\kappa +2)+\alpha _2 \kappa +\alpha _2}{g \kappa },\nonumber\\
\epsilon_4&=& \frac{\sqrt{\alpha _1^2 \kappa ^2+\alpha _2^2 (\kappa -1)^2}+\alpha _1 (\kappa +2)+\alpha _2 \kappa +\alpha _2}{g \kappa }\nonumber,
\end{eqnarray}
and the respective non-normalized eigenstates 
\begin{eqnarray}
\vert \chi_1\rangle&=&\left(1,\frac{\sqrt{\alpha _1^2 \kappa ^2+\alpha _2^2}+\alpha _2}{\alpha _1 \kappa },\frac{\sqrt{\alpha _1^2 \kappa ^2+\alpha _2^2}+\alpha _2}{\alpha _1 \kappa },1\right),\\
\vert \chi_2\rangle&=&\left(1,\frac{\alpha _2-\sqrt{\alpha _1^2 \kappa ^2+\alpha _2^2}}{\alpha _1 \kappa },\frac{\alpha _2-\sqrt{\alpha _1^2 \kappa ^2+\alpha _2^2}}{\alpha _1 \kappa },1\right),\nonumber\\
\vert \chi_3\rangle&=&\left(-1,\frac{\alpha _2 (\kappa -1)-\sqrt{\alpha _1^2 \kappa ^2+\alpha _2^2 (\kappa -1)^2}}{\alpha _1 \kappa },\frac{\sqrt{\alpha _1^2 \kappa ^2+\alpha _2^2 (\kappa -1)^2}-\alpha _2 \kappa +\alpha _2}{\alpha _1 \kappa },1\right),\nonumber\\
\vert \chi_4\rangle&=&\left(-1,\frac{\sqrt{\alpha _1^2 \kappa ^2+\alpha _2^2 (\kappa -1)^2}+\alpha _2 (\kappa -1)}{\alpha _1 \kappa },\frac{-\sqrt{\alpha _1^2 \kappa ^2+\alpha _2^2 (\kappa -1)^2}-\alpha _2 \kappa +\alpha _2}{\alpha _1 \kappa },1\right)\nonumber.
\end{eqnarray}
The eigenvalues and eigenstates presented above can be compared to the ones in \cite{artem2} by taking $\alpha_i/g=J_i$. We have also omitted the constant energy term $E_0$ in these expressions. In the regime of $\kappa<1$, the two lowest energy states are given by $\epsilon_3$ (ground state) and $\epsilon_1$ (first excited state). We can therefore write an analytical expression for the frequency $\omega_J$ that defines the tunneling of the impurity between the wells. It is given by

\begin{equation}
\omega_J=\frac{\alpha_2\kappa-\sqrt{\alpha_1^2 \kappa^2+\alpha_2^2}+\sqrt{\alpha_1^2 \kappa^2+\alpha_2^2 (\kappa-1)^2}}{g \kappa},
\end{equation}
where again we considered $\alpha_3=\alpha_1$. In Fig.\ref{figap4} we show the behavior of this frequency with increasing $\kappa$ in the case of $b=0$. The result shown here can be directly related to the line defined by $b=0$ in Fig. 4 of the main text. For small $\kappa$, we get $\omega_J=\frac{\alpha_1^2 \kappa^2}{2 g \alpha_2}$, while for $\kappa\gg 1$ (approaching the fermionic limit for the background gas) it becomes constant: $\omega_J=\frac{\alpha_2-\alpha_1+\sqrt{\alpha_1^2+\alpha_2^2}}{g}$.

\section*{References}

\bibliographystyle{ieeetr}
\bibliography{biblio}

\end{document}